\documentstyle[12pt]{article}

  \advance\oddsidemargin  by -1in
  \advance\evensidemargin by -1in
  \advance\topmargin      by -0.5in
\newcommand{\Z}{{\bf Z}}

\newcommand{\R}{{\bf R}}
\newcommand{\C}{{\bf C}}
\newcommand{\T}{{\bf T}}
\newcommand{\B}{{\bf B}}
\newcommand{\LL}{{\bf L}}
\newcommand{\U}{{\bf U}}
\newcommand{\D}{{\bf D}}

\newcommand{\size}{\mathop{\it size}\nolimits}
\newcommand{\poly}{\mathop{\it poly}\nolimits}
\newcommand{\St}{\mathop{\rm St}\nolimits}
\newcommand{\Tr}{\mathop{\rm Tr}\nolimits}
\newcommand{\tr}{\mathop{\rm tr}\nolimits}
\newcommand{\Prob}{\mathop{\rm Prob}\nolimits}
\newcommand{\const}{\mathop{\rm const}\nolimits}

\newtheorem{definition}{\bf Definition}
\newtheorem{theorem}{\bf Theorem}
\newtheorem{lemma}{\bf Lemma}
\def\trlist{\trivlist \def\makelabel##1{\hskip\labelsep##1}}
\newenvironment{proof} {\trlist \item[{\bf Proof.}]}
                       {\qquad\unskip$\Box$\endtrivlist}

\newenvironment{remark} [1] {\trlist \item[{\bf #1.}]} {\endtrivlist}

\newcommand{\calB}{{\cal B}}
\newcommand{\calC}{{\cal C}}
\newcommand{\calR}{{\cal R}}
\newcommand{\calQ}{{\cal Q}}
\newcommand{\calM}{{\cal M}}
\newcommand{\calN}{{\cal N}}
\newcommand{\calV}{{\cal V}}
\newcommand{\calW}{{\cal W}}
\newcommand{\calE}{{\cal E}}

\title{Quantum measurements and the Abelian Stabilizer Problem}

\author{ {\bf A.~Yu.~Kitaev}\medskip\\
{\it L.D.Landau Institute for Theoretical Physics,} \\
{\it Kosygina St.~2, Moscow 117940, Russia}\smallskip \\
e-mail:\quad kitaev\,@\,itp.ac.ru
\vspace{1cm} }

\begin{document}

\maketitle

\begin{abstract}
We present a polynomial quantum algorithm for the Abelian stabilizer problem
which includes both factoring and the discrete logarithm. Thus we extend famous
Shor's results~\cite{Shor}. Our method is based on a procedure for
measuring an eigenvalue of a unitary operator. Another application of this
procedure is a polynomial quantum Fourier transform algorithm for an arbitrary
finite Abelian group. The paper also contains a rather detailed introduction
to the theory of quantum computation.
\end{abstract}

\section*{Introduction}

It has been known for long time that all ``reasonable'' computation models
are equivalent. Moreover, every universal machine $A$ can simulate any other
machine $B$ with at most polynomial slowdown. For instance, a computation,
which takes time $t$ on a random access memory (RAM) machine, can be done in
time $O(t^2)$ on a Turing machine. (The slowdown is nonlinear because the
Turing machine has to scroll its tape to access a distant memory cell).
In view of this equivalence, theoretical computer scientists classify
algorithms as polynomial\footnote{
 An algorithm (for a given problem and a given machine model) is called
 polynomial if the number of steps of the algorithm grows not faster than some
 power of the size of the input.}
and superpolynomial, the former being considered efficient, the latter
inefficient. A polynomial algorithm remains polynomial when adapted to another
machine model.

Many physical phenomena can be simulated on a computer in polynomial time,
although it is sometimes impracticable because of the great number of particles
involved. However, simulation of quantum mechanics may be computationally
expensive even with few particles.
Consider a system with $2$ states. If we take $n$ copies of this system we
will get a new system with $2^n$ states. Its quantum evolution (for a given
time interval) is characterized by a unitary matrix of size $2^n\times 2^n$.
Unless one invents a more intelligent method, simulation of the evolution
amounts to multiplication of evolution matrices corresponding to very short
time intervals. It takes exponential time to compute one separate item of the
product. (However, such computation can be done with polynomial memory).

But if quantum mechanics is really difficult to simulate, a quantum computer
should be more powerful than the classical one. How to know it for certain?
A quantum computer is still an imaginary device which has not been
constructed yet. Not thinking about technology, there are 3 fundamental
questions to be answered.
\begin{enumerate}
\item Is there any simple and universal model of quantum computation?
\item Can a quantum computer solve a computational problem which is known to be
hard for a classical computer?
\item As far as the group of unitary transformations $\U(2^n)$ is continuous:
To what extent is quantum computation sensitive to perturbation? And is it
possible to organize computation so that a moderate perturbation would not
affect the result?
\end{enumerate}

Quantum devices for doing classical computation were suggested by Benioff~
\cite{Benioff}, Peres~\cite{Peres} and Feynmann~\cite{Feynmann}.
Deutsch~\cite{Deutsch1,Deutsch2} was the first to give an explicit model of
quantum computation. He defined both quantum Turing machines and quantum
circuits. Yao~\cite{Yao} showed that these two models are equivalent.
More specifically, quantum Turing machines can simulate, and be simulated by,
uniform families of polynomial size quantum circuits, with at most polynomial
slowdown. Quantum circuits are generally more convenient for developing
quantum algorithms.

Quantum circuits are rather generic quantum systems which can simulate other
quantum systems. We have seen that such simulation may be problematic with a
classical computer, so the answer to the second question is probably ``yes''.
However, we do not know whether simulating quantum mechanics on a classical
computer is really hard. In fact, if no efficient  algorithm is known for a
problem, it doesn't mean that such an algorithm doesn't exist.
Unfortunately, no reasonable computational problem has been {\em proven} to be
hard yet. So it is interesting to find efficient quantum algorithms
for problems which are considered as hard by computer science experts.
The most remarkable result of this type has been obtained
by Shor~\cite{Shor} who invented polynomial quantum algorithms for the discrete
logarithm and factoring of integers. However, it is not clear yet whether a
polynomial quantum algorithm exists for an NP-complete problem.

In order to obtain a correct result under perturbation,
every step of the computation must be done with precision\,
$c\,(\mbox{number of steps})^{-1}$\, (the constant $c$ depends on the allowed
error probability, see Sec.~\ref{sec_quco}). Thus the number of precision
bits, needed to specify each elementary quantum operator (gate), is
logarithmic~\cite{Bernstein-Vazirani}. This precision requirement is rather
weak, which gives hope that quantum computation can be done by a physical
device. Note that exponential precision (i.e. polynomial number of precision
bits) is almost certainly infeasible; fortunately, it is not needed for
quantum computation. However, even polynomial precision may prove to be
impractical. A fully satisfactory solution would be to do arbitrarily long
computation with fixed gate precision, by use of some error correction
procedure. Alternatively, one should ensure high precision by some physical
mechanism beyond the formal computation model. Precision still remains the
most important problem in the field of quantum computation.

In this paper we suggest a polynomial quantum algorithm for a so-called Abelian
Stabilizer Problem (ASP) which includes both factoring and the discrete
logarithm. Thus we reproduce Shor's result by a different method. Another
special case of the ASP was studied by Grigoriev~\cite{Grigoriev} in
connection with the shift equivalence problem for polynomials. The ASP should
have some applications to computational problems in number theory and
algebraic geometry, but this topic needs a separate study.

The key point of our solution is a concept of quantum measurement. We also use
a generalization of Simon's procedure~\cite{Simon} for finding a certain group
of characters. In Sec.~\ref{sec_qurevers} we demonstrate a
more subtle use of quantum measurements by describing a polynomial algorithm
for the Quantum Fourier Transform (QFT) on an arbitrary finite Abelian group.
This doesn't solve any classical computational problem because the QFT is
defined in terms of quantum mechanics. However, the construction itself may be
interesting.
Polynomial QFT algorithms were known for groups $(\Z_2)^k$ \cite{Deutsch-Jozsa}
and $\Z_{q}$, where $q=2^n$ \cite{Coppersmith} or $q$ is a smooth number,
i.e. contains no prime power factor larger than $(\log q)^c$ \cite{Shor}.

\section{The Abelian Stabilizer Problem}\label{sec_asp}

Let $G$ be a group acting on a finite set $M$. Suppose that this action and the
group operations in $G$ can be computed easily. Compute the stabilizer of a
given element $a\in M$. This problem (still to be formulated in a more rigorous
language) includes many interesting cases, e.g. graph isomorphism.
Unfortunately, we are not able now to treat the problem in its generality.
Rather, we will assume that the group $G$ is Abelian.
As far as any finitely generated Abelian group is a homomorphic image of
$\Z^k$, we may set w.~l.~o.~g. $G=\Z^k$.~\footnote{
  For the group $G=(\Z_{p})^k$, Grigoriev~\cite{Grigoriev} designed a quantum
  algorithm which was polynomial in $k$ and $p$ (but not in $\log p$).}
We will also assume that the set $M$ can be identified,
by some one-to-one coding, with a subset of a Boolean cube $\B^n=\{0,1\}^n$.
(Our algorithm does not work if each element of $M$ have many
representations in $\B^n$, even if the equivalence of these representations
can be checked by an efficient procedure). This restricted problem is called
the Abelian Stabilizer Problem (ASP). We proceed with an exact definition.

An ASP (more exactly, an instance of the ASP) consists of the following items:
\begin{itemize}
\item Two positive integers $k$ and $n$. The pair $(k,n)$ is called the size
of the problem.
\item An element $a\in\B^n$.
\item A function $F:\Z^k\times M\to M$\,\,\, ($a\in M\subseteq\B^n$),\,\,
such that
\begin{displaymath}
  F(0,x)\,=\,x \qquad F(g+h,x)\,=\,F(g,F(h,x)) \qquad
  \mbox{for any}\quad g,h\in\Z^k,\quad x\in M
\end{displaymath}
\end{itemize}
The function $F$ should be regarded as a blackbox subroutine which receives an
input $(g,x)\in\Z^k\times\B^n$\, and produces an output $y\in\B^n$,\,
so that $y=F(g,x)$ for every\, $g\in\Z^k$,\,\,$x\in M$. (If $x\not\in M$, the
subroutine may fail or give an arbitrary result. We do not assume that the
condition $x\in M$ is checkable). This subroutine $F$ can be invoked by a
quantum computer in the way precisely defined in Sec.~\ref{sec_compmod}.

\begin{remark}{Remark}
In all reasonable applications (see examples below) the function $F$ can be
computed
in polynomial time. A quantum computer can do this job itself, so there is no
need to use a blackbox subroutine in this case. Let us describe this situation
more exactly. Denote by $\size(g)$ the number of bits needed to represent
an element $g\in\Z^k$ (in a reasonable coding).\footnote{
  In different reasonable codings $\size(g)$ may differ at most by a constant
  factor.}
Let $\poly$ stand for any function that grows not faster than a polynomial,
i.e. $\poly(x)=x^{O(1)}$. Suppose that the subroutine $F$ is a classical or
even quantum machine (see Sec.~\ref{sec_compmod} for explicit models) which
computes $F(g,x)$ in time $\poly(\size(g)+n)$ at most. With a fixed function
$\poly$, this defines a restricted class of ASPs. In this case we will get a
polynomial quantum algorithm which uses a {\em description} of the machine $F$
rather than invokes it as a subroutine.
\end{remark}

The stabilizer of $a$ with respect to $F$ is the set\,
$\St_F(a)=\{g\in\Z^k:\, F(g,a)=a\}$.\,
This is a subgroup in $\Z^k$ of index~ $\le|M|\le 2^n$. Hence $\St_F(a)$ is
isomorphic to $\Z^k$ and has a basis $(g_1,\ldots,g_k)$ of polynomial size,\,
meaning that $\sum_{j=1}^{k}\size(g_j)\le\poly(n+k)$. Any such basis is
acceptable as a solution of the ASP. There is an efficient procedure which
checks whether $(g_1,\ldots,g_k)$ and $(g'_1,\ldots,g'_k)$ represent the same
subgroup in $\Z^k$.
Given a subgroup $A\subseteq\Z^k$ of rank $k$ represented by a polynomial size
basis, one can compute (by a very simple polynomial algorithm)
a unique {\em canonical basis} $(h_1,\ldots,h_k)$.
This basis is given by the columns of the matrix $(m_{ij},\ i,j=1,\ldots,k)$\,
uniquely characterized by the conditions
\begin{equation}
  \begin{array}{rcllc}
    & m_{ij} & \!=\, 0             \quad & \mbox{if} & i>j \\
    & m_{ii} & \!>\, 0 & & \\
    0\,\le\! & m_{ij} & <\, m_{ii} \quad & \mbox{if} & i<j
  \end{array}
\end{equation}
Thus finding an arbitrary polynomial size basis for the stabilizer is
equivalent to finding the canonical one.

Factoring and the discrete logarithm can be reduced to the ASP.
Let $M$ be the ring
of integers modulo $q$,\,\, $G$ the group of invertible elements of $M$.
If $g_1,\ldots,g_k\in G$ then\,
$F_{g_1,\ldots,g_k}: (m_1,\ldots,m_k,x)\mapsto g_1^{m_1},\ldots,g_k^{m_k}x$\,\,
($m_i\in\Z$,\,\, $x\in M$)\, is an action of $\Z^k$ on $M$. Consider two cases.
\begin{enumerate}
\item {\bf Factoring.}\, The stabilizer of $1$ with respect to $F_g$ gives the
order of an element $g$ in the group $G$. There is a randomized reduction from
factoring to the order of an element~\cite{factoring}. (A sketch of this
reduction can be found in Shor's paper~\cite{Shor}).
\item {\bf Discrete logarithm.}\, Let $q$ be a prime,\,
$\zeta\in G\cong\Z_{q-1}$ a primitive element,\, $g\in G$ an arbitrary element.
The stabilizer of $1$ with respect to $F_{\zeta,g}$ is\,
$P=\{(m,r)\in\Z^2:\, \zeta^m g^r=1\}$.\, Given a basis of the subgroup
$P\subseteq\Z^2$, we can find an element of the form $(m,-1)\in P$. Then
$\zeta^m=g$.
\end{enumerate}

\section{Computation models}\label{sec_compmod}

This section is intended mostly for a reader not familiar with the
subject. We define the models usually used in the field of quantum computation.
A more experienced reader should just pay attention to a few non-common terms
and notations.

In Sec.~\ref{sec_circ} we define Boolean circuits and operation sequences.
These two models are trivially equivalent. The language of operation
sequences is not quite common but we find it convenient. It is closer to an
intuitive model of computation and allow simpler notations. (Circuits can be
nicely represented by diagrams, but we do not use diagrams in this paper).
We also briefly discuss the concept of uniformity.

In Sec.~\ref{sec_revers} we overview the concept of reversible computation
introduced by Lecerf~\cite{Lecerf} and Bennett~\cite{Bennett}.
This is an important link between the standard models (e.g. Boolean circuits)
and quantum computation. The results of this section have quantum analogues
(see Sec.~\ref{sec_qurevers}).

In Sec.~\ref{sec_quform} we summarize the basic concepts and notations of
quantum mechanics.

In Sec.~\ref{sec_quco} we give a formal model of quantum computation and
discuss its basic properties.

\subsection{Boolean circuits and operation sequences}\label{sec_circ}

 From now on, we often use functions of type $f:\B^n\to\B^m$.
We write $n=\delta(f)$,\,\, $m=\rho(f)$.\,

Let $\calB$ be a set of such functions to be used as elementary blocks for
building more complicated functions. The set $\calB$ is called a {\em basis};
its elements are called {\em gates}. Usually one uses the standard basis
$\calC=\{\neg,\wedge\}$ (negation and the ``and'' function).
This basis is {\em complete}, that is any
Boolean function can be represented as a composition of the basis elements.

Let $F:\B^n\to\B^m$ be an arbitrary function.
A {\em Boolean circuit} for $F$ is a procedure which converts an input
$x\in\B^n$ to the output $y=F(x)\in\B^m$\,
working with auxiliary Boolean variables $z_{1},\ldots,z_{K}$ according
to the following instructions:
\begin{enumerate}
\item Copy $x$ to ($z_{1},\ldots,z_{n})$.
\item Compute $z_{n+1},\ldots,z_{K}$ in sequel, using some gates
$f_{i}\in\calB$\,\, ($i=1,\ldots,L$)\, and variables already computed.
More specifically,
\begin{equation}
  \Bigl( z_{k_{i}+1},\ldots,z_{k_{i}+\rho(f_{i})} \Bigr)\ =\
  f_{i} \Bigl( z_{\alpha(i,1)},\ldots,z_{\alpha(i,\delta(f_{i}))} \Bigr)
  \qquad\quad
  \alpha(i,1),\ldots,\alpha(i,\delta(f_{i}))\ \le\ k_{i} \quad
\end{equation}
where\,\,
$k_{1}=n$,\,\,\, $k_{i+1}=k_{i}+\rho(f_{i})$.
\item Read $y$ from ($z_{\beta(1)},\ldots,z_{\beta(m)})$.
\end{enumerate}
Thus a Boolean circuit is defined by the sequence of functions
$f_{1},\ldots,f_{L}\in\calB$\, and the numbers $\alpha(i,j)$,\, $\beta(j)$.
The number $L$ is called the {\em size} of the circuit. For more generality,
we may assume $F$ to be a {\em partial function} $\B^n\to\B^m$, that is a
function $N\to\B^m$, where $N\subseteq\B^n$. In this case the output $y$ must
coincide with $F(x)$ for every $x\in N$.

\begin{remark}{Circuits and algorithms}
A Boolean circuit can only work with inputs of fixed size. However, computation
problems are usually defined for inputs of variable size ---
consider, for example, the number addition problem $(x,x')\mapsto x+x'\,$.
Any reasonable computational problem can be represented by a family of
functions $F|_s:\B^{s}\to\B^{\poly(s)}$, each corresponding to a particular
input size $s$. One needs a separate Boolean circuit for each $F|_s$. If a
polynomial algorithm exists for $F$ then each function $F|_s$ can be computed
by a circuit ${\cal F}_{s}$ of size $\poly(s)$. In fact, a computer
(e.g. a Turing machine), working in\,
$\mbox{space}\times\mbox{time}\le t\times t$,\,
can be simulated by a Boolean circuit of size $O(t^2)$.

However, the existence of a polynomial
size circuit ${\cal F}_{s}$ for each $F|_s$ does not necessarily imply that the
total function $F$ can be efficiently computed. For this, one must be able to
construct the circuits ${\cal F}_{s}$ efficiently. More exactly, the function
$s\mapsto{\cal F}_{s}$ must be computable on a Turing machine in polynomial
time. A family of circuits (${\cal F}_{s}$), which satisfies this condition,
is called {\em uniform}. Thus the machine produces a circuit, and the circuit
computes the function.

This two-level construction is especially good for defining non-standard
computation models, including quantum computation. In Sec.~\ref{sec_quco} we
will define some theoretical quantum devices which can compute Boolean
functions. Although these devices operate in a quantum way, they allow
classical description, i.e. each particular device can be represented by a
binary word. By a quantum algorithm for a problem $F$ we will mean a
{\em classical} algorithm which constructs a quantum device $\Phi_{s}$ for
each function $F|_s$.
\end{remark}

In a Boolean circuit the value of each variables $z_i$ is computed only once.
However, in real computers memory cells can be reused to operate with
new information. At each step the computer does some operation with a few
memory cells. Let us give a simple model of such computation.

Denote by $\Delta=\{1,\ldots,K\}$ the memory to be used in computation.
Each memory element (bit) $i\in\Delta$ represents a Boolean variable $z_{i}$.
Any ordered collection of bits is called a {\em register}.
Associated with a register $A=(A_1,\ldots,A_n)$
is the variable $z_{A}=(z_{A_1},\ldots,z_{A_n})$ taking values from $\B^n$.
(Here $A_i\in\Delta$\, and $A_{i}\not=A_{j}$ if $i\not=j$).
Given a Boolean operator $g:\B^n\to\B^n$,\, we can define its action
$g[A]:z_{A}\mapsto g(z_{A})$ on the set of states of the register $A$. (By a
{\em Boolean operator} we mean an arbitrary mapping of a Boolean cube into
itself). We may regard $g[A]$ as an operator on the total set of memory
states,\, $\Gamma=\B^\Delta$.

Now take some set (basis) $\cal B$ of Boolean operators.
Operators of the form $g[A]$\,\, ($g\in\calB$)\, will be called
{\em operations}. The new model is a procedure of the following type:
\begin{enumerate}
\item Place the input into some register $X$.
      Set all the other bits equal to $0$.
\item Do some operations $g_{1}[A_{1}],\ldots,g_{L}[A_{L}]$\,\,
($g_i\in\calB$)\, one by one.
\item Read the output from some register $Y$.
\end{enumerate}

A Boolean circuit may be considered as the sequence of operators
\begin{equation} \label{circsim}
  g_{i}:\ \Bigl( u_{1},\ldots,u_{\delta(f_i)},\ v_{1},\ldots,v_{\rho(f_i)}
  \Bigr)\, \mapsto\,
  \Bigl( u_{1},\ldots,u_{\delta(f_i)},\ f_i(u_{1},\ldots,u_{\delta(f_i)})
  \Bigr)
\end{equation}
applied to the registers\,
$A_{i}=\Bigl(\alpha(i,1),\ldots,\alpha(i,\delta(f_{i})),\,
 k_{i}+1,\ldots,k_{i}+\rho(f_i)\Bigr)$.\,
On the other hand, any sequence of $L$ operation can be simulated by a Boolean
circuit of size $L$ --- one should just reserve a separate variable for each
new Boolean value that appears during computation.
So these two models are equivalent.

\subsection{Reversible computation}\label{sec_revers}

The models defined above, as well as operation of a real computer, are not
reversible. In fact, even erasing a bit (i.e. setting it equal to $0$) is not
reversible. However, the laws of quantum mechanics are reversible, since the
inverse of a unitary matrix exists and is also a unitary matrix. So, before
passing on to quantum computation, one must be able to do classical
computation reversibly.

Of course, reversible computation must use only bijective gates
$g_i:\B^n\to\B^n$, i.e. permutation on Boolean cubes. A simple but
important example is the bijective
operator $\tau_n:(u,v)\mapsto(u,\,v\oplus u)$\, on $\B^{2n}$,\,
where ``$\oplus$'' stands for the bitwise addition modulo~2.
(Obviously, applying $\tau_n$ is the same as to apply the operator
$\tau=\tau_1$ to each pair of bits).
The operator $\tau_n$ allows to copy the content of one register into another,
provided the second register is empty.
For a more general example, consider an arbitrary function $F:N\to\B^m$\,\,
($N\subseteq\B^n$),\,\, then
\begin{equation}
  F_{\tau}:\, N\times\B^m\to N\times\B^m\ :\qquad
  (u,v)\,\mapsto\,(u,\,v\oplus F(u))
\end{equation}
is a bijection. It is quite clear now how to simulate a Boolean
circuit by a sequence of bijective operations. Instead of the operators~
(\ref{circsim}) one should take the operators $(f_i)_{\tau}$. The result will
be the same because
$v=(v_{1},\ldots,v_{\rho(f_i)})=(z_{k_i+1},\ldots,z_{k_i+\rho(f_i)})$
is zero before the operator $(f_i)_{\tau}$ is applied.

Formally, this observation is enough to proceed with quantum computation.
However, the above computation with bijective gates is not truly reversible.
In fact, besides the output it produces some ``garbage'', i.e. extra
information which have to be forgotten after the computation is finished.
Without this garbage the computer cannot run back from the output to the input.
We will see that garbage does not allow to use the result of a computation in
an essentially quantum way. It is worth noting that a real computer also
produces some sort of garbage, namely heat. (Actually, the existing computers
produce much more heat than necessary). It is rather surprising that the
garbage in our model can be avoided.

First of all, we are to give an exact definition of computation without
garbage, usually called {\em reversible computation}. In what follows we
assume the memory $\Delta$ to be the union of two disjoint registers, an
input-output register $X$ and an auxiliary register $W$. Thus a state of the
memory is denoted as $(x,w)$,\, where\, $x\in\B^X$,\,\, $w\in\B^W$.

\begin{definition}\label{def_revers}
Let $G:N\to M$\,\, ($N,M\subseteq\B^n$)\, be an arbitrary bijection.
A sequence of bijective operations $g_{i}[A_{i}]$\,\, ($i=1,\ldots,L$)\, is
said to {\em represent} $G$, or compute $G$ reversibly,\, if their
composition\, $g_{L}[A_{L}]\circ\ldots\circ g_{1}[A_{1}]$\,
maps $(x,0)$ to $(G(x),0)$ for every $x\in N$.
\end{definition}

\begin{lemma}\label{lem_garbrem1}
Suppose that a function $F:N\to\B^m$\,\, ($N\subseteq\B^n$)\, is computable
in a basis $\calB$ by a Boolean circuit of size $L$. Then $F_\tau$ can be
represented in the basis\, $\calB_\tau=\,\{f_\tau:f\in\calB\}\cup\{\tau\}$\,
by an operation sequence of length\, $2L+m$.
\end{lemma}

\begin{proof}
The Boolean circuit can be simulated by a sequence of $L$ operations from the
basis $\calB_\tau$. We may assume that this simulation uses
registers $U$, $W$ and $Y$\, for input, intermediate results and output,
respectively,\, where $U\cap W=\emptyset$ and $Y\subseteq U\cup W$.
The total effect of the simulation can be represented by an operator
$G=G[U,W]$. Let $V$ be a new register of size $m$. Then $X=U\cup V$ can be
used as an input-output register for reversible computation of the function
$F_\tau$. We can denote a memory state as $(u,v,w)$, where\, $u$, $v$ and $w$
stand for the contents of $U$, $V$, and $W$, respectively.
The needed reversible computation is given by the operator\,
\begin{displaymath}
  (G[U,W])^{-1} \,\circ\, \tau_m[Y,V] \,\circ\, G[U,W]\ :\qquad
  (u,v,0) \,\mapsto\, (u,\,v\oplus F(u),\, 0) \qquad
  ( u\in N,\,\ v\in\B^m )
\end{displaymath}
Indeed, the operator $G[U,W]$ computes $F(x)$,\, the operator $\tau_m[Y,V]$
adds it to $v$ modulo~2,\, and ($G[U,W])^{-1}$ removes the garbage, that is
makes $w$ equal to $0$.
\end{proof}

\begin{lemma}\label{lem_garbrem2}
Let $G:N\to M$\,\, ($N,M\subseteq\B^n$) be a bijection.
Suppose that $G$ and $G^{-1}$ are computable in a basis $\calB$
by Boolean circuits of size $L$ and $L'$, respectively. Then $G$ can be
represented in the basis\, $\calB_\tau$\, by an operation sequence of length
$2L+2L'+4n$.
\end{lemma}

\begin{proof}
Let $X$ be the input-output register, $Y$ an auxiliary register of the same
size $n$. We should add also another auxiliary register $W$ to be used
implicitly in the reversible subroutines $G_\tau$ and $(G^{-1})_\tau$. By the
previous lemma,
these subroutines need $2L+n$ and $2L'+n$ operations, respectively.
The required computation is given by the operator
\begin{displaymath}
  \tau_n[X,Y] \,\circ\, \tau_n[Y,X] \,\circ\,
  (G^{-1})_\tau[Y,X] \,\circ\, G_\tau[X,Y]
\end{displaymath}
Indeed,\,\,
$(x,0) \,\mapsto\, (x,G(x)) \,\mapsto\, (0,G(x)) \,\mapsto\, (G(x),G(x))
 \,\mapsto\, (G(x),0)$.
\end{proof}

\begin{remark}{Corollaries}\mbox{} \samepage
\begin{enumerate}
\item Any permutation of $n$ bits can be done by $4n$ operations $\tau$.
\item The basis $\calC_\tau$ is complete for reversible computation.
\end{enumerate}
\end{remark}

The gate $\neg_\tau\in\calC_\tau$ may be replaced with $\neg$. Thus we get
another complete basis\, $\calR = \{\neg,\tau,\wedge_\tau\}$.~\footnote{
  The gate $\tau$ can be represented in terms of $\wedge_\tau$ and $\neg$,
  so it is not necessary.}
We will always use this basis unless we speak about so-called relative
computation, that is {\em computation with a blackbox subroutine}.
\begin{definition}
Let\, $F:\B^n\to\B^m$\, be an arbitrary function, possibly partial.
Reversible computation in the basis\, $\calR\cup\{F_\tau\}$\, is called
{\em reversible computation with subroutine $F$}.
\end{definition}
This definition is natural due to Lemma~\ref{lem_garbrem1}. Actually, we will
need only one particular case of a blackbox subroutine.

Let $F:\Z^k\times M\to\B^n$ be the function from the definition of an ASP. It
is not a Boolean function, so the above definition should be modified. Let
\begin{equation}
{\cal Z}^k_s\ =\ \Bigl\{ g\in\Z^k:\, \size(g),\size(-g)\le s\Bigr \}
\end{equation}
We can identify ${\cal Z}^k_s$ with a certain subset of $\B^s$.
Denote by $F|_{s+n}$ the restriction of $F$ to ${\cal Z}^k_s\times M$.
By computation with the subroutine $F$ we will mean computation with
$F|_{s+n}$,\, where $s=\poly(k+n)$. Our quantum algorithm will use the
following bijection
\begin{equation}\label{G}
  G:\, \Z^k\times M\, \to\, \Z^k\times M\ :\qquad
  (g,x)\, \mapsto\, (g,F(g,x))
\end{equation}
Note that\, $G^{-1}:(g,x)\mapsto(g,F(-g,x))$.\, The function $G|_{s+n}$, the
restriction of $G$, may be considered as a partial bijective operator on
$\B^s\times\B^n$. By Lemma~\ref{lem_garbrem2}, this function can be easily
computed with the subroutine $F|_{s+n}$.

\subsection{The quantum formalism}\label{sec_quform}

In this subsection we remind the reader the quantum formalism for a system
with a finite set of states $\Gamma$. In the computation-theoretic context,\,
$\Gamma=\B^\Delta$ is the set of states of a computer memory $\Delta$.

A {\em quantum state} is characterized by a unit vector $|\psi\rangle$
in the complex space $\C(\Gamma)=\C^\Gamma$ equipped with a Hermitian scalar
product $\langle\cdot|\cdot\rangle$. To be exact, the term ``quantum state''
is usually used to denote a one-dimensional subspace of $\C(\Gamma)$, i.e. a
unit vector up to a phase factor $e^{i\phi}$.~\footnote{
This definition is motivated by the fact that the
probability~(\protect\ref{quprob}) is invariant under the transformation
$|\psi\rangle\mapsto e^{i\phi}|\psi\rangle$,\,
so the phase factor may be neglected in many cases.}
Corresponding to the {\em classical states}
$a\in\Gamma$ are the {\em standard vectors} $|a\rangle\in\C(\Gamma)$ which form
an orthonormal basis of $\C(\Gamma)$. Time evolution of a quantum system is
given by a transformation of the form\, $|\psi\rangle\mapsto U|\psi\rangle$,\,
where $U$ is a unitary operator.
Any bijection $G:\Gamma\to\Gamma$ may be regarded
as a unitary operator acting by the rule\, $G|a\rangle=|G(a)\rangle$.\,
Such operators are called {\em classical}.

Elements of $\C(\Gamma)$ are usually denoted like $|\xi\rangle$,
even if the symbol in the brackets is never used alone. The scalar
product of two vectors $|\xi\rangle,|\eta\rangle\in\C(\Gamma)$\, is denoted by
$\langle\xi|\eta\rangle$. Thus $\langle\xi|$ stands for the linear functional\,
$|\eta\rangle\mapsto\langle\xi|\eta\rangle$\, on $\C(\Gamma)$.
The space of such functionals is denoted by $\C(\Gamma)^*$.
If\, $|\xi\rangle=\sum_{j\in\Gamma}c_{j}|j\rangle$\, then\,
$\langle\xi|=\sum_{j\in\Gamma}c_{j}^*\langle j|$.\, In the coordinate
representation
\begin{displaymath}
  |\xi\rangle\ =\ \left( \begin{array}{l} c_{1}\\ \vdots\\ c_{k} \end{array}
                  \right) \qquad\qquad
  \langle\xi|\ =\ \Bigl(\, c_{1}^*,\, \ldots,\, c_{k}^* \,\Bigr)
\end{displaymath}
If $|\xi\rangle,|\eta\rangle\in\C(\Gamma)$\, then $|\xi\rangle\langle\eta|$ is
an element of $\C(\Gamma)\otimes\C(\Gamma)^*$ and thus may be considered as a
linear operator on $\C(\Gamma)$. The result of the application of a linear
operator $A:\C(\Gamma)\to\C(\Gamma)$ to a vector $|\xi\rangle$ is denoted by\,
$A|\xi\rangle=|A\xi\rangle$. Thus
\begin{displaymath}
  \langle\xi|A\eta\rangle\, =\, \langle\xi|A|\eta\rangle\, =\,
  \langle A^\dagger\xi|\eta\rangle \qquad\qquad\quad
  \langle\xi|A\, =\,  \langle A^\dagger\xi|
\end{displaymath}
where $A^\dagger$ is the operator adjoint to $A$.

The algebra of linear operators $\C(\Gamma)\to\C(\Gamma)$ is denoted by
$\LL(\Gamma)$,\, while $\U(\Gamma)$ denotes the group of unitary operators.

Let $\Pi_\calM|\xi\rangle$ denote the orthogonal projection of a vector
$|\xi\rangle$ onto a linear subspace $\calM\subseteq\C(\Gamma)$. The projection
operator $\Pi_\calM$ can be represented as
$\sum_{j=1}^{k} |e_j\rangle\langle e_j|$,\,
where\, $(|e_j\rangle,\ j=1,\ldots,k)$\, is an arbitrary orthonormal basis
of $\calM$.

Two things are most important in the quantum formalism: the probabilistic
interpretation of quantum mechanics and the relation between a system and its
subsystems. From the mathematical point of view, the probabilistic
interpretation is just a definition of some function called ``probability''.
After the definition
is given, one can check that this function does have some basic properties of
classical probability. Here we just give the definition. The analogy with the
classical case will be fully developed in the Sec.~\ref{sec_qumeas} where we
introduce conditional probabilities.

The classical probability\, $P(\mu,M)=\mu(M)=\sum_{j\in M}\mu(j)$\, is a
function of two arguments: a probability measure $\mu$ on $\Gamma$ and a subset
$M\subseteq\Gamma$. (As far as the set $\Gamma$ is finite, a probability
measure is simply a positive function $\mu:\Gamma\to\R$,\, such that\,
$\sum_{j\in\Gamma}\mu(j)=1$).\, Correspondingly, the quantum probability
depends on a quantum state $|\xi\rangle$ and a linear subspace
$\calM\subseteq\C(\Gamma)$
\begin{equation}\label{quprob}
  P(\xi,\calM)\ =\ \langle\xi|\Pi_\calM|\xi\rangle
\end{equation}
This quantity can be also represented as $\Tr(\rho\Pi_\calM)$,\, where
$\rho=|\xi\rangle\langle\xi|$ is the {\em density operator} associated
with the state $|\xi\rangle$. In a more general setting, a density operator on
$\Gamma$ is an arbitrary positive Hermitian operator $\rho\in\LL(\Gamma)$
with trace $1$;\, the set of such operators is denoted by $\D(\Gamma)$.
In this case we write
\begin{equation}\label{quprob1}
P(\rho,\calM) =\ \Tr(\rho\Pi_\calM)
\end{equation}
This definition includes the classical probability. Indeed, let $\calM$ be the
subspace generated by the standard vectors $|a\rangle:a\in M$. Let also
$\rho=\sum_{a\in\Gamma}\mu(a)|a\rangle\langle a|$,\, where $\mu$ is a
probability measure on $\Gamma$. Then $P(\rho,\calM)=P(\mu,M)$.
Like the classical probability, the quantum probability is additive.
Specifically, if $\calM$ and $\calN$ are orthogonal subspaces then\,
$P(\rho,\,\calM\oplus\calN)=P(\rho,\calM)+P(\rho,\calN)$.\,
A generic density operator is said to represent a {\em mixed state} of the
system, while quantum states defined above are called {\em pure}.
Time evolution of a density operator is given by the formula\,\,
$\rho\mapsto U\rho U^\dagger$.

Let our system consist of two subsystems, $A$ and $B$,\, that is
$\Gamma=\Gamma_A\times\Gamma_B$,\, where $\Gamma_A$ and $\Gamma_B$ are the
classical state sets of the subsystems. Two vectors\,
$|\xi_A\rangle\in\C(\Gamma_A)$,\,\,\, $|\xi_B\rangle\in\C(\Gamma_B)$\,
can be combined to give the vector
\begin{displaymath}
  |\xi_A,\xi_B\rangle\, =\, |\xi_A\rangle\otimes|\xi_B\rangle\ \in\
  \C(\Gamma_A)\otimes \C(\Gamma_B)\, =\, \C(\Gamma)
\end{displaymath}
One can also define tensor product of linear subspaces, linear
operators, unitary operators and density operators. It is clear that
\begin{equation}\label{indep}
  P(\rho_A\otimes\rho_B,\,\calM_A\otimes\calM_B)\ =\
  P(\rho_A,\calM_A)\, P(\rho_B,\calM_B)
\end{equation}
The most striking difference between quantum mechanics and classical
mechanics is that a quantum state of a whole system can not be generally
decomposed into states of subsystems. In fact, one can not even define any
natural linear mapping $\C(\Gamma)\to\C(\Gamma_A)$. (That is the reason why
we have to avoid garbage in computation, see an explanation below).
However, a density operator $\rho$ on $\Gamma$ can be ``projected'' onto
$\Gamma_A$ to give the density operator
\begin{equation}
  \rho_A\ =\ \Tr_B\rho\ =\ \sum_{a,b\in\Gamma_A}\ |a\rangle\,
  \left( \sum_{c\in\Gamma_B} \langle a,c|\rho|b,c\rangle \right)\, \langle b|
\end{equation}
One may pass on to the projection and consider its evolution separately
if the subsystem $A$ does not interact with $B$ in future. Indeed,
\begin{displaymath}
  \Tr_B\Bigl( (U_A\otimes U_B)\,\rho\, (U_A\otimes U_B)^\dagger \Bigr)\ =\
  U_A\, (\Tr_B\rho)\, U_A^\dagger  \qquad\qquad
  P\Bigl( \rho,\, \calM_A\otimes\C(\Gamma_B) \Bigr)\ =\
  P(\Tr_B\rho,\,\calM_A)
\end{displaymath}
Note that the projection of a pure state is generally a mixed state.

Finally, let us introduce a concept of a quantum variable, or
{\em observable}.\footnote{
Our definition of an observable slightly differs from the conventional one.}
Let $\Omega$ be a family of mutually orthogonal linear subspaces of
$\C(\Gamma)$. Denote by $\calV_{?}$ the orthogonal complement to
$\bigoplus_{\calV\in\Omega}\calV$. In this setting, we say
that an observable $z_\Omega$ is defined. Let\, $\rho\in\D(\Gamma)$,\,\,
$\calV\in\Omega$.\, Then the quantity $P(\rho,\calV)$ is called
{\em the probability of $z_\Omega$ to have the value $\calV$}.
Obviously,\, $\sum_{\calV\in\Omega}P(\rho,\calV)\,+P(\rho,\calV_?)\,=1$.\,
Thus $P(\rho,\calV_?)$ is the probability that $z_\Omega$ has no value.
If $\cal A$ is a predicate on $\Omega$
(i.e. a function $\Omega\to\{\mbox{true},\mbox{false}\}$)\,
then $\Prob_\rho[{\cal A}(z_\Omega)]$\, denotes the
probability of ${\cal A}(z_\Omega)$ being true
\begin{displaymath}
  \Prob_\rho[{\cal A}(z_\Omega)]\ =\
  \sum_{\calV\in\Omega:\,{\cal A}(\calV)} P(\rho,\calV)
\end{displaymath}
For example,\, $\Prob_\rho[z_\Omega=\calV]\,=\,P(\rho,\calV)$.\,
The notation $\Prob_\rho[\ldots]$ is convenient because it expresses the
intuitive meaning of probability.

\subsection{Quantum computation}\label{sec_quco}

\subsubsection{The basic model}

Before giving a formal model of quantum computation, we will describe
elementary operations with a quantum system which seem feasible from the
physical point of view. From now on, we assume that
$\Gamma=\B^\Delta$,\, where $\Delta=\{1,\ldots,K\}$ is a memory used in
computation.

Let $\Delta=A\cup B$,\, where $A$ and $B$ are two disjoint
registers. Thus $\Gamma=\Gamma_A\times\Gamma_B$,\, where $\Gamma_A$ and
$\Gamma_B$ are the sets of states of the registers $A$ and $B$.
Let $U\in\U(\B^n)$, where $n=|A|$. As far as the Boolean cube $\B^n$ can be
identified with $\Gamma_A$,\, we can define
the action $U[A]$ of the operator $U$ on the space $\C(\Gamma_A)$. By tensoring
with the unit operator $1[B]\in\U(\Gamma_B)$,\, we can make $U[A]$ to be an
operator on the space $\C(\Gamma)$ corresponding to the whole system.
Physical implementation of such an operator seems feasible provided the number
$n$ is small.

For each $a\in\Gamma_A$\, denote by $\calW_a$ the subspace
$(|a\rangle)\otimes\C(\Gamma_B)\subseteq\C(\Gamma)$.
These subspaces are mutually orthogonal.
Thus the {\em standard observable} $z_A$ associated with the register
$A$ is defined. It always have some value $a\in\B^n$, meaning that
$\bigoplus_{a\in\B^n}\calW_a=\C(\Gamma)$. Given a quantum state
$|\xi\rangle$, it is possible to measure the value of the observable $z_A$,
that is to organize some physical procedure which gives a result $a$ with
probability $P(\xi,\calW_a)$. For this, it is enough to measure the state of
the whole memory\, (the result $c\in\Gamma$ is obtained with probability\,
$P(\xi,c)=|\langle c|\xi\rangle|^2\,$)\,
and then ignore information contained in the register $B$.
(Certainly, this works for mixed states as well). The measurement destroys
the quantum state, so it must be done in the end of computation.

We are going to define a quantum model which is similar to general
(i.e. garbage-producing) sequences of bijective operations.
(Reversible quantum computation will be considered later on).
We assume that computer memory $\Delta$ is a disjoint union of the input
register $X$ and an auxiliary register $W$,
the output register $Y\subseteq\Delta$ being arbitrary\,\,
($|X|=n$,\,\, $|Y|=m$).\,
Thus a classical state of the memory can be denoted as
$(x,w)$,\, where\, $x\in\B^X=\B^n$,\,\, $w\in\B^W$.

\begin{definition}\label{def_quco}
Let $\calB$ be a basis of unitary operators,\,\,
$0<\epsilon<\frac{1}{2}$\, an arbitrary constant.
A sequence of operations\, $U_{1}[A_{1}],\ldots,U_{L}[A_{L}]$\,\,
($U_i\in\calB$)\, is said {\em to compute a function\, $F:N\to\B^{m}$\,\,
($N\subseteq\B^n$)\, with error probability $\,\le\epsilon$}\, if
\begin{displaymath}
  \forall\, x\in N\quad\ \Prob\,_{U|x,0\rangle}\,\Bigl[ z_{Y}=F(x) \Bigr]\
  \ge\ 1-\epsilon \qquad\quad \mbox{where}\quad
  U\,=\,U_{L}[A_{L}]\ldots U_{1}[A_{1}]
\end{displaymath}
\end{definition}

The error probability can be made arbitrary small by repeating the computation
several times. Indeed, let us take $k$ different copies of the memory and do
the same computation in each of them independently, with the same input
$x\in N$. Due to~(\ref{indep}), the corresponding outputs $y_1,\ldots,y_k$ may
be considered as independent random variables. By definition, the eventual
result is $y$ if more than a half of all $y_i$ are equal to $y$.
The total probability of an error or failure does not exceed\,
$\sum_{j\ge k/2}{k\choose j}\epsilon^{j}(1-\epsilon)^{k-j}\,\le\,
 \lambda^k$,\,\,
where $\lambda=2\,(\epsilon(1-\epsilon))^{1/2}<1$.\, Within the scope of
polynomial computation, the error probability can be made as small as\,
$\exp(-\poly(n))$,\, where $\poly$ is an arbitrary function of polynomial
growth. Note that the original choice of the constant $\epsilon$ is not
important; one usually sets $\epsilon=\frac{1}{3}$.

\noindent{\bf Remark.\,}
The above procedure can be represented by the formula\,
$y\ =\ \mbox{MAJ}(y_1,\ldots,y_k)$,\,
where $\mbox{MAJ}$ is a partial function called the {\em majority function}.
To make it work, one must be able to compute this function in the basis
$\calB$. This is possible, for example, in the classical basis $\calR$.

\begin{remark}{The choice of the basis}
In this paper we use the basis\, $\calQ=\U(\B^1)\cup\{\tau,\wedge_\tau\}$.\,
Note that $\neg\in\U(\B^1)$,\, so $\calR\subseteq\calQ$.
Hence any classical reversible computation can be done in the basis $\calQ$.
Actually, this basis is complete for quantum computation; even its proper
subset $\U(\B^1)\cup\{\tau\}$ is a complete basis~\cite{9authors}.
If a blackbox subroutine $F$ is given, we add the operator $F_\tau$
to the basis.\footnote{
  If the $F$ is a partial function, the operator $F_\tau$ is partial.
  In general, a partial unitary operator is a bijective norm-preserving linear
  operator between two subspaces.}
There is still one problem with our choice: the basis $\calQ$ is infinite so
infinite information is needed to specify its element. Fortunately, quantum
computation can be done with polynomial gate precision (see below). Hence
logarithmic number of precision bits is sufficient.
\end{remark}

\subsubsection{Precision}

Precision of a vector $|\xi\rangle\in\C(\Gamma)$ can be characterized by means
of the usual (Hermitian) norm\,
$\|\xi\rangle\|=\sqrt{\langle\xi|\xi\rangle}$.\,
There are two natural norms on the space of linear operators $\LL(\Gamma)$,\,
the usual operator norm
\begin{displaymath}
  \|A\|\ =\ \sup_{|\xi\rangle\not=0} \frac{\|A|\xi\rangle\|}{\|\xi\rangle\|}
\end{displaymath}
and the trace norm
\begin{displaymath}
  \|A\|_{\tr}\ =\ \Tr\sqrt{A^\dagger A}\ =\,\
  \inf\left\{ \sum_{j} \|\xi_j\rangle\|\,\|\langle\eta_j\|\,:\
              \sum_{j} |\xi_j\rangle\langle\eta_j| = A \right\}\,\ =\,\
  \sup_{B\not=0}  \frac{|\Tr AB|}{\|B\|}
\end{displaymath}
The most important properties of these norms are as follows
\begin{equation}\label{normprop}
  \|AB\|\, \le\, \|A\|\,\|B\| \qquad\quad
  \|AB\|_{\tr},\, \|BA\|_{\tr}\ \le\ \|B\|\,\|A\|_{\tr} \qquad\quad
  |\Tr A|\, \le\, \|A\|_{\tr}
\end{equation}
We say that a unitary operator $\tilde U$ {\em represents a unitary operator
$U$ with precision $\delta$}\, if\, $\|\tilde U-U\|\le\delta$.\,
The following lemma shows that errors are simply added through computation
but are not amplified.

\begin{lemma}\label{lem_accum}
Let\, $U_1,\ldots,U_L$, $\tilde U_1,\ldots,\tilde U_L$\, be unitary operators.
If\, $\tilde U_j$ represents\, $U_j$ with precision $\delta_j$
for $j=1,\ldots,L$\,\,
then\, $\tilde U_L\ldots\tilde U_1$\, represents\, $U_L\ldots U_1$\, with
precision\, $\delta_1+\ldots+\delta_L$.
\end{lemma}
\begin{proof}
If $L=2$ then\,
$\|\tilde U_2\tilde U_1-U_2 U_1\| \,\le\,
 \|(\tilde U_2-U_2)\tilde U_1\| \,+\, \|U_2(\tilde U_1-U_1)\| \,\le\,
 \|(\tilde U_2-U_2)\| + \|\tilde U_1-U_1\|$,\, since a unitary operator has
the norm~$1$. The general case follows by induction.
\end{proof}

The trace norm
$\|\cdot\|_{\tr}$ is suitable to characterize precision of density operators.
Note that if $|\xi\rangle,|\eta\rangle$ are unit vectors then
\begin{equation}\label{densdiff}
  \Bigl\| |\xi\rangle\langle\xi| - |\eta\rangle\langle\eta| \Bigl\|_{\tr}\ =\
  2\,\sqrt{1-|\langle\xi|\eta\rangle|^2}\ \le\
  2\,\Bigl\| |\xi\rangle - |\eta\rangle \Bigr\|
\end{equation}

\begin{lemma}\label{lem_probdiff}
Let $\Omega$ be a family of mutually orthogonal linear subspaces of
$\C(\Gamma)$. Then for any pair of density operators $\rho$, $\gamma$
\begin{displaymath}
  \sum_{\calV\in\Omega} |P(\rho,\calV)-P(\gamma,\calV)|\ \le\
  \|\rho-\gamma\|_{\tr}
\end{displaymath}
\end{lemma}
\begin{proof}
The left hand side of this inequality can be represented as
$\Tr((\rho-\gamma)B)$,\, where\, $B=\sum_{\calV\in\Omega}(\pm\Pi_\calV)$.\,
It is clear that $\|B\|\le 1$. Then use the norm properties~(\ref{normprop}).
\end{proof}

Combining Lemma~\ref{lem_accum} with the inequality~(\ref{densdiff}) and
Lemma~\ref{lem_probdiff}, we obtain the following
\begin{lemma}\label{lem_precision}
Let an operation sequence of length $L$ compute a function with error
probability $\,\le\epsilon$. If each operation is represented with precision
$\delta$ then the resulting error probability does not exceed\,
$\epsilon+2L\delta$.
\end{lemma}
Thus the necessary gate precision is $\const L^{-1}$. Note that classical
(non-reversible) computation can be simulated without error accumulation,
by use of error correcting codes.

The notion of precision is also applicable to partial operators. Let $U$ and
$\tilde U$ be {\em partial unitary operators on $\C(\Gamma)$}. In other
words,\, $U:\calN\to\calM$ and $\tilde U:\tilde\calN\to\tilde\calM$ are
bijective linear
operators preserving the scalar product, where\,
$\calN,\calM,\tilde\calN,\tilde\calM\subseteq\C(\Gamma)$.\,
We say that the operator $\tilde U$ represents $U$ with precision $\delta$\,
if\, $\calM\subseteq\tilde\calM$,\,\, $\calN\subseteq\tilde\calN$\, and\,
$\|(U-A)|\xi\rangle\|\le\delta\|\xi\rangle\|$\, for any $|\xi\rangle\in\calN$.
Note that $\tilde U^{-1}$ represents $U^{-1}$ with the same precision.
Lemma~\ref{lem_accum} remains valid for partial unitary operators.

\subsubsection{Reversible quantum computation}

Definition~\ref{def_revers} can be extended to the quantum case in a
straightforward way. One can also define approximate reversible computation.
In view of the above consideration, it is convenient to use the language of
partial operators. The set of partial unitary operators on $\C(\Gamma)$ will
be denoted by $\check\U(\Gamma)$. Denote by $\omega$ the partial operator
$|0\rangle\mapsto|0\rangle$ on $\C(\B^k)$ (for any $k$).
Let the memory $\Delta$ be the union of two disjoint
registers, an input-output register $X$ and an auxiliary register $W$.
A state of the memory is denoted as $(x,w)$,\, where\,
$x\in\B^X$,\,\, $w\in\B^W$.
\begin{definition}
Let $U\in\check\U(\B^n)$.\,
A sequence of operations\, $U_{1}[A_{1}],\ldots,U_{L}[A_{L}]$\, is said to
{\em represent} $U$ (with precision $\delta$)\,
if the operator\, $U_{L}[A_{L}]\ldots U_{L}[A_{L}]$\, represents the partial
operator\, $U[X]\otimes\omega[W]$\, (with precision $\delta$).
\end{definition}

As in the classical case, a non-reversible quantum computation procedure can
be converted  into a reversible one (see Sec.~\ref{sec_qurevers} for more
detail).

\subsubsection{Quantum gates with control parameters}

Let $U:\calN\to\calM$\,\, ($\calN,\calM\subseteq\C(\B^n)$)\, be a linear
operator. Define a new operator\,
$\Lambda(U):\C(\B^1)\otimes\calN\to\C(\B^1)\otimes\calM$\, by the formula
\begin{equation}
  \Lambda(U)\, |a,\xi\rangle\ =\ \left\{
  \begin{array}{ll}
    |0,\xi\rangle \quad & \mbox{if}\ a=0 \\
    |1\rangle\otimes\,U|\xi\rangle \quad & \mbox{if}\ a=1
  \end{array} \right.
\end{equation}
Thus the operator $U$ is applied or not depending on whether an additional
{\em control bit}\,\footnote{
  Note that in our model the control bit is quantum, as all the other bits.}
is equal to $1$ or $0$. For example,\, $\Lambda(\neg)=\tau$,\,\,
$\Lambda(\tau)=\wedge_\tau$.\, Another example:
\begin{displaymath}
  \Lambda(e^{i\phi})\ =\
  \left( \begin{array}{cc} 1&0\\ 0&e^{i\phi} \end{array} \right) \qquad\qquad
  (\phi\in\R)
\end{displaymath}
(The number $e^{i\phi}$ can be considered as a unitary operator on $\C(\B^0)$).
It is obvious that
\begin{equation}
  \Lambda(UV)\ =\ \Lambda(U)\,\Lambda(V) \qquad\qquad
  \Lambda(V^{-1}UV)[1,A]\,\ =\,\ V^{-1}[A]\ \Lambda(U)[1,A]\ V[A]
\end{equation}

For a classical operator $U$, the operator $\Lambda(U)$ can be computed by
a Boolean circuit in the basis $\calC\cup\{U\}$. Hence it can be represented in
the basis $\calR\cup\{U,U^{-1}\}$\, (by Lemma~\ref{lem_garbrem2}). This does
not work in the general case. However, the following statement holds.
\begin{lemma}\label{lem_control}
Let $U$ be a (partial) unitary operator on $\C(\B^n)$,\, such that
$U|0\rangle=|0\rangle$.\, Then the operator $\Lambda(U)$ can be represented
in the basis $\calQ\cup\{U\}$ by an operation sequence of length $4n+1$,
the gate $U$ being used only once.
\end{lemma}
\begin{proof} Let the input-output register be $X=\{1\}\cup A$,\, where $1$
denotes the control bit. Let $B$ be an auxiliary register of size $n$.
The required computation is given by the composition of operators
\begin{displaymath}
  \Lambda(\tau_n)[1,A,B]\quad \Lambda(\tau_n)[1,B,A]\quad U[B]\quad
  \Lambda(\tau_n)[1,B,A]\quad \Lambda(\tau_n)[1,A,B]
\end{displaymath}
\end{proof}
\begin{remark}{Corollary}
For any $U\in\U(\B^1)$ the operator $\Lambda(U)$ can be represented
in the basis $\calQ$.\\
Indeed, $U$ can be represented as $V^{-1}WVe^{i\phi}$,\,
where $W|0\rangle=|0\rangle$.
\end{remark}

Let us also consider a more general type of control. For any function
${\cal U}:\B^l\to\check\U(\B^n)$ we define the operator
\begin{equation}
  \Lambda({\cal U})\, \in\, \check\U(\B^l\times\B^n) \qquad\qquad
  \Lambda({\cal U})\, |a,\xi\rangle\ =\
  |a\rangle\otimes\, {\cal U}(a)|\xi\rangle
\end{equation}
\begin{lemma}\label{lem_controlchange}
Let $F:\B^k\to\B^l$ be a partial function;\, ${\cal U}:\B^l\to\check\U(\B^n)$.
Consider two operators,\, $T=\Lambda({\cal U})\in\check\U(\B^l\times\B^n)$\,
and\,  $F_T=\Lambda({\cal U}\circ F)\in\check\U(\B^k\times\B^n)$.\,
If the function $F$ can be computed by a Boolean circuit of size $L$ in a basis
$\cal B$\, then the operator $F_T$ can be represented by an operation sequence
of length $2L+1$ in the basis $\calB_\tau\cup\{T\}$,\,
the gate $T$ being used only once.
\end{lemma}
(Proof is quite similar to the proof of Lemma~\ref{lem_garbrem1}).
\medskip

As an application of this lemma, we will show how to create an arbitrary
unit vector\, $|\eta\rangle=u|0\rangle+v|1\rangle\in\C(\B^1)$\, if $u$ and $v$
are given as control parameters. For simplicity, assume that $u,v\in\R$, that
is\, $u=\cos\theta$,\,\, $v=\sin\theta$. Then the vector $|\theta,\eta\rangle$
can be obtained from $|\theta,0\rangle$ by applying the operator
$R:\,|\theta,\xi\rangle\,\mapsto\,|\theta\rangle\otimes R_\theta|\xi\rangle$.
Here $\theta$ is a real number represented, with some precision,
in a binary form;
\begin{displaymath}
  R_\theta\ =\ \left(
  \begin{array}{rr} \cos\theta & -\sin\theta \\ \sin\theta & \cos\theta
  \end{array} \right)
\end{displaymath}
Lemma~\ref{lem_controlchange} allows to construct the operator $R$ from
$\Lambda(R_\theta)$ with $\theta=2\pi\,2^{-s}$\,\, ($s=1,2,\ldots$).

\subsubsection{Some other properties of quantum computation}

\begin{remark}{Simulating classical probability}
To simulate classical probabilistic
computation, one needs to create random bits.
Let us take a quantum bit in the state\,
$2^{-1/2}\,\Bigl(|0\rangle+|1\rangle\Bigr)$\,
and copy it to another bit by the operator $\tau$. (Beware that the
operator $\tau$ copies each classical state entering a quantum superposition,
not the whole superposition!) Thus we get the two-bit quantum state\,
$|\psi\rangle=\, 2^{-1/2}\,\Bigl(|0,0\rangle+|1,1\rangle\Bigr)$.\,
Then discard the copy (or just not use it in computation). This situation can
be described by transition to a density operator corresponding to the first
bit only
\begin{displaymath}
  \rho_1\ =\ \Tr_2\Bigl( |\psi\rangle\langle\psi| \Bigr)\ =\
  \left( \begin{array}{cc} 1/2&0\\ 0&1/2 \end{array} \right)
\end{displaymath}
This density operator corresponds to the classical probability measure\,
$\mu(0)=\mu(1)=\frac{1}{2}$.
\end{remark}

\begin{remark}{The effect of garbage}
Let $G:\B^n\to\B^n$\, be a classical operator to be used in quantum
computation. Assume that the operator $G$ is computed by a sequence of
bijective operations. We are to show that the operator $G$ must be computed
without garbage, otherwise quantum coherence will be destroyed.
Suppose that garbage is produced. Then $G$ is actually represented by an
operator $U:(x,0)\mapsto(G(x),g(x))$\, on the total set of memory states.
The operator $U$ transforms a quantum state
$|\xi\rangle=\sum_{x}c_{x}|x\rangle$\, into the state\,
$|\psi\rangle=\,\sum_{x}\,c_{x}\,|G(x),g(x)\rangle$.\, As far as the garbage
is ignored, we should take the trace with respect to the second variable.
Thus we get the density operator
\begin{displaymath}
\rho\ =\ \sum_{x,y:\, g(x)=g(y)}\, c_{x}^*c_{y}\, |G(x)\rangle\langle G(y)|
\end{displaymath}
If the garbage $g(x)$ is the same for all $x$ then\,
$\rho=G|\xi\rangle\langle\xi|G^\dagger$,\, so the operator $U$ does what it is
supposed to do. Now consider the worst case: different inputs produce different
garbage. Then the density operator\,
$\rho=\,\sum_{x}\,|c_{x}|^2\, |G(x)\rangle\langle G(x)|$\,
is classical;\, it could be obtained if we first measured the value of $x$ and
then applied $G$ in a classical way. We conclude that a classical operator can
not be used in an essentially quantum way unless it is computed reversibly.
\end{remark}

\section{Quantum measurements}\label{sec_qumeas}

One of the physical assumptions, underlying the formal model of quantum
computation, is the possibility to measure the classical state of the memory.
Such measurement is a specific type of interaction between the quantum computer
and an external physical device. Description of the measurement procedure is
beyond the scope of our formal analysis.
However, we can formally define and study another
type of measurement in which one part of the computer works as a device
measuring the state of another part. We will see that such measurement obeys
the usual laws of conditional probability. So, if subsystems $A_1,A_2,\ldots$
measure each other in sequence, this process can be simulated by a Markov
chain. This fact is very important for understanding the probabilistic
interpretation of quantum mechanics in physical context.
We may believe that the chain of measurements
extends beyond the system in study, and the last measurement done by an
external device is of the same type. Except for this philosophical remark,
we will use quantum measurements as a concrete tool for developing quantum
algorithms.

\begin{definition}
Let $A$ and $D$ be two disjoint registers,\, $\Omega$ a family of mutually
orthogonal subspaces of $\C(\B^A)$.\,
Set\, $\calN=\bigoplus_{\calV\in\Omega}\calV$.
\begin{enumerate}
\item A {\em measurement operator} for the observable $z_\Omega$\,
is a linear operator of the form
\begin{displaymath}
  U\ =\ \sum_{\calV\in\Omega} \Pi_{\calV}\otimes U_{\calV}\quad : \qquad
  \calN\otimes\C(\B^D)\,\to\,\calN\otimes\C(\B^D)
\end{displaymath}
where $U_{\calV}$ are arbitrary unitary operators on $\C(\B^D)$.
\item A measurement operator $U$ together with a register $C\subseteq D$\,
is called a {\em measurement} with result $z_C$. Let $|C|=m$.
Denote by $\calW_y$ the subspace of $\C(\B^D)$ corresponding to the
situation $z_C=y$,\, i.e.\,
$\calW_y=(|y\rangle)\otimes\C(\B^{D\backslash C})$.\,
The numbers
\begin{displaymath}
  P_{U,C}(\calV,y)\ =\ P\Bigl( U_{\calV}|0\rangle,\,\calW_{y} \Bigr)
  \qquad\qquad (\calV\in\Omega,\,\ y\in\B^m)
\end{displaymath}
are called the {\em conditional probabilities} for the measurement $(U,C)$.
\item A measurement $(U,C)$ is said to {\em measure} the value of a function
$F:\Omega\to\B^{m}$ with error probability $\,\le\epsilon$\,\,
if\,\, $P_{U,C}(\calV,F(\calV))\ge 1-\epsilon$\,\,
for every $\calV\in\Omega$.
\end{enumerate}
\end{definition}

For example, the operator $\tau_n[A,D]$ is a measurement for the observable
$z_A$. Any quantum computation (see Definition~\ref{def_quco}) can be
organized as a measurement with respect to its input. For this, it suffice to
copy the input by the operator $\tau_n$ and use the copy instead of the
original. Alternatively, one can use the bits of the input as control
parameters, e.g. in operators $\Lambda(U)$.

\begin{remark}{Important example}
Let $U$ be a unitary operator on a subspace $\calN\subseteq\C(\B^n)$.
The eigenvalues of this operator have the form\,
$\lambda(\phi)=\exp(2\pi i\phi)$,\,
where $\phi$ is a real number $(\bmod\ 1)$. Denote by $\calE(U,\phi)$ the
corresponding eigenspaces. Without risk of confusion, the corresponding
observable may be denoted simply by $\phi$.

Let the operator $U$ act on a register $A$. Denote by $1$ an additional bit
and introduce the matrix
\begin{displaymath}
  S\ =\ \frac{1}{\sqrt{2}} \left( \begin{array}{rr} 1&1\\
        1&-1 \end{array} \right)
\end{displaymath}
Then the operator
\begin{equation}
  \Xi(U)[A,1]\,\ =\,\ S[1]\,\ \Lambda(U)[1,A]\,\ S[1]
\end{equation}
is a measurement operator for the observable $\phi$.
If $|\xi\rangle\in\calE(U,\phi)$\, then\,
$\Xi(U)|\xi,0\rangle=|\xi,\eta\rangle$,\, where
\begin{displaymath}
  |\eta\rangle\,\ =\quad \frac{1}{2}\
  \left( \begin{array}{rr} 1&1\\ 1&-1 \end{array} \right)\
  \left( \begin{array}{cc} 1&0\\ 0&\lambda(\phi)  \end{array} \right)\
  \left( \begin{array}{rr} 1&1\\ 1&-1 \end{array} \right)\
  \left( \begin{array}{c} 1\\ 0 \end{array} \right)\quad =\quad
  \left( \begin{array}{l} \frac{1}{2}(1+\lambda(\phi))\smallskip\\
                          \frac{1}{2}(1-\lambda(\phi))
         \end{array} \right)
\end{displaymath}
Hence the conditional probabilities\,
$P_{\Xi(U)}(\phi,y)\,=\,P_{\,\Xi(U)[A,1],\ 1\,}(\calE(U,\phi),y)$\,
are as follows
\begin{equation}
  P_{\Xi(U)}(\phi,0)\ =\ \frac{1}{2}\Bigl( 1+\cos(2\pi\phi) \Bigr)
  \qquad\qquad
  P_{\Xi(U)}(\phi,1)\ =\ \frac{1}{2}\Bigl( 1-\cos(2\pi\phi) \Bigr)
\end{equation}
\end{remark}

General properties of measurement operators and measurements are quite simple.
Let us fix a register $A$ and a family $\Omega$ of mutually orthogonal
subspaces in $\C(\B^A)$. Set\, $\calN=\bigoplus_{\calV\in\Omega}\calV$.\,
We will consider measurement operators for the same
observable $z_\Omega$ with different additional registers $D$.
\begin{lemma}\label{lem_meas}
\mbox{}\samepage
\begin{enumerate}
\item Let $(U,C)$ be a measurement with an additional register $D\supseteq C$.
Then for any quantum state $|\xi\rangle\in\calN$\,
the composite probability formula holds
\begin{equation}\label{comprob}
  \Prob_{\,U|\xi,0\rangle\,}\Bigl[ z_C=y \Bigr]\,\ =\,\
  \sum_{\calV\in\Omega}\, P(|\xi\rangle,\calV)\ P_{U,C}(\calV,y)
\end{equation}
\item The product of several measurements operators is a measurement operator.
Measurement operators with disjoint additional registers commute.
\item Let\, $(U',C')$ and $(U'',C'')$ be measurements with disjoint additional
registers $D'\supseteq C'$ and $D''\supseteq C''$.\,\,
Set\,\, $U=U'U''$,\,\,\,  $C=C'\cup C''$.\,\, Then
\begin{equation}\label{indepcond}
  P_{U,C}\Bigl( \calV,\,(y',y'') \Bigr)\ =\
  P_{U',C'}(\calV,y')\ P_{U'',C''}(\calV,y'')
\end{equation}
\end{enumerate}
\end{lemma}
\begin{proof}\samepage
\item[{\bf 1.\,}] It is clear that\,
$U|\xi,0\rangle=\,\sum_{\calV\in\Omega}\Pi_{\calV}\,U|\xi,0\rangle$.\, Hence
\begin{displaymath}
  \Prob\,_{U|\xi,0\rangle}\,\Bigl[ z_C=y \Bigr]\,\ =\,\
  \sum_{\calV\in\Omega}\,
  \langle\xi,0|U^\dagger\, \Pi_{\calV}\Pi_{\calW_y}\,  U|\xi,0\rangle\,\ =\,\
  \sum_{\calV\in\Omega}\, \langle\xi|\Pi_{\calV}|\xi\rangle\
  \langle 0|U_{\calV}^\dagger\, \Pi_{\calW_y} U_{\calV}|0\rangle
\end{displaymath}
This gives the right hand side of eq.~(\ref{comprob}).
\item[{\bf 2.\,}] It follows from the definition.
\item[{\bf 3.\,}] It follows from the general property of quantum
probability~(\ref{indep}).\\*
\end{proof}

Now let us return to the example considered above. Suppose that the measurement
operator $\Xi(U)$ is applied $s$ times to the same register $A$ and different
additional bits $1,\ldots,s$. Then one can measure the values $z_1,\ldots,z_s$
of these bits and count how many $1$'s are contained in the resulting sequence
of $0$'s and $1$'s. Denote this count by $y$. Thus a new measurement
$\Xi_{s}(U)$ is defined, the number $y$ being its result.
Since $z_1,\ldots,z_s$ behave as independent random variables,
$y/s$ is most likely close to $P_{\Xi(U)}(\phi,1)$.
More exactly, for any given constant $\delta>0$
\begin{displaymath}
  \Prob\Bigl[\, |y/s-P_{\Xi(U)}(\phi,1)|\,>\,\delta \,\Bigl]\,\ \le\,\
  2\,\exp(-c(\delta)\,s)
\end{displaymath}
where $c(\delta)>0$. Thus we measure the quantity\,
$P_{\Xi(U)}(\phi,1)=\frac{1}{2}\Bigl(1-\cos(2\pi\phi)\Bigr)$\, with precision
$\delta$ and error probability $\,\le\exp(-c(\delta)\,s)$. If we substitute
$iU$ for $U$ then $\cos(2\pi\phi)$ will change to $-\sin(2\pi\phi)$. So we can
measure both $\cos(2\pi\phi)$ and $\sin(2\pi\phi)$; this information is enough
to find $\phi$. We have proved the following

\begin{lemma}\label{lem_egvalapprox}
Let $\delta>0$ be a constant. For any $\epsilon>0$,\, the value of the
observable $\phi$ can be measured with precision $\delta$ and error
probability $\le\epsilon$\, by an operation sequence of length
$O(\log(1/\epsilon))$ in the basis $\calQ\cup\{\Lambda(U)\}$.
\end{lemma}

Unfortunately, it is difficult to measure $\phi$ with arbitrary precision
because the cost of measurement (i.e. the length of the operation sequence)
grows polynomially in $\delta$. However, the situation is different if we have
in our disposal the operators $\Lambda(U^k)$ for all $k$. More specifically,
consider the operator
\begin{equation}
  U^{[0,r]}\,:\ \C(\{0,\ldots,r\})\otimes\calN\,\to\,
  \C(\{0,\ldots,r\})\otimes\calN\, \qquad\qquad
  U^{[0,r]}\,|a,\xi\rangle\ =\ |a\rangle\otimes U^{a}|\xi\rangle
\end{equation}
where $r=2^l-1$. (Note that the set $\{0,\ldots,2^l-1\}$
can be naturally identified with $\B^l$). By Lemma~\ref{lem_controlchange},
the operators\, $\Lambda\!\left(U^{2^j}\right)$\,\, ($j=0,\ldots,l-1$)\,
can be represented in terms of $U^{[0,r]}$.
It takes $O(\log(l/\epsilon))$ operation to localize each of the numbers\,\,
$2^j\phi\ (\bmod\ 1)$\,\, in one of the $8$ intervals\,
$\left[\frac{s-1}{8},\,\frac{s+1}{8}\right]$\,\, ($s=0,\ldots,7$)\,\, with
error probability $\le\epsilon/l$. Using this information, one can find
(by a polynomial algorithm) the value of $\phi$ with precision\,
$\frac{1}{8}\,2^{-(l-1)}=2^{-l-2}$\,
and error probability $\le\epsilon$. We have obtained the following result

\begin{lemma}\label{lem_egvalapprox1}
Let $l$ be a positive integer;\, $r=2^l-1$.\,
For any $\epsilon>0$,\, the value of the observable $\phi$ can be measured
with precision $2^{-l-2}$ and error probability $\le\epsilon$\, by an
operation sequence of length\, $O(l\,\log(l/\epsilon))+\poly(l)$\,
in the basis $\calQ\cup\{U^{[0,r]}\}$. (The gate $U^{[0,r]}$ is used at most
$O(l\,\log(l/\epsilon))$ times).
\end{lemma}

Now consider an important particular case: $U$ is a permutation on a subset
$N\subseteq\B^n$. Corresponding to each cycle of the permutation are
eigenvalues of the form $\exp(2\pi i\frac{p}{q})$,\, where $q$ is the length
of the cycle. Hence the values of $\phi$ are rational numbers with
denominators $\le 2^n$.
The minimal separation between such numbers is $\Bigl(2^n(2^n-1)\Bigr)^{-1}$.
Consequently, the exact value of $\phi$ can be found by measuring it with
precision $2^{-2n-1}$. Moreover, the transition from the measured value to the
exact one can be performed in polynomial time, using continuous fractions.
What follows is a brief proof of this claim.

Suppose that the measurement produced a number\, $\phi'=p'/q'$\,\,
($0\le p'<q'=2^{2n+1}$),\,\, such that\,
$|\phi'-\phi|\le 2^{-2n-1}\ (\bmod\ 1)$.\,
It is easy to check whether $\phi=0$, so we will assume that $\phi\not=0$.
Thus $\phi=p/q$, where $p$ and $q$ are mutually prime,\, $0<p<q\le 2^n$.
Let us define a sequence of positive integers $(k_1,\ldots,k_s)$ which can be
obtained by applying Euclid's algorithm to the pair $(q,p)$
\begin{displaymath}
  q_{j-1}\ =\ k_{j}q_{j}+q_{j+1}\,, \qquad 0\,\le\, q_{j+1}\,<\, q_{j}
  \qquad\quad (j=1,\ldots,s)
\end{displaymath}\nopagebreak
\begin{displaymath}
  q_{0}\,=\, q \qquad q_{1}\,=\, p \qquad\quad
  q_{s}\,=\, \mbox{g.c.d.}(q,p)\,=\, 1 \qquad q_{s+1}\,=\, 0
\end{displaymath}
Since $q$ and $p$ are not known, we can not compute $k_1,\ldots,k_s$ directly.
Instead of that, we can apply Euclid's algorithm to the pair $(q',p')$ to get
a sequence $(k'_1,\ldots,k'_{s'})$. It is easy to show that
\begin{displaymath}
  \frac{q_{j}}{q_{j-1}}\,-\,\frac{1}{2q_{j-1}^2}\,\ \le\,\
  \frac{q'_{j-1}}{q'_{j}}\,\ < \,\
  \frac{q_{j}}{q_{j-1}}\,+\,\frac{1}{(2q_{j-1}-1/q_{j})\,q_{j-1}} \qquad\qquad
  (j=1,\ldots,s)
\end{displaymath}\nopagebreak
\begin{displaymath}
  k'_{j}\ =\ k_{j} \qquad (j=1,\ldots,s-1) \qquad\qquad\
  k'_{s}\,=\,k_{s}\quad\ \mbox{or}\quad\quad
  k'_{s}=k_{s}-1,\quad k'_{s+1}=1
\end{displaymath}
It follows that\, $p/q=\mbox{CF}(0,k'_{1},\ldots,k'_{s})$\, or\,
$p/q=\mbox{CF}(0,k'_{1},\ldots,k'_{s+1})$,\, where
\begin{displaymath}
  \mbox{CF}(m)\ =\ m, \qquad\quad
  \mbox{CF}(m_0,m_1,\ldots)\ =\ m_0\,+\,\frac{1}{\mbox{CF}(m_1,\ldots)}
\end{displaymath}
To find $\phi=p/q$,\, we can compute the numbers\,
$\phi_j=\mbox{CF}(0,k'_{1},\ldots,k'_{j})$\,\, ($j=1,2\ldots$)\, until\,
$|\phi'-\phi_j|\le 2^{-2n-1}$. Then $\phi=\phi_j$. We have proved the following
\begin{theorem}\label{th_meas}
Let $U$ be a permutation on a set $N\subseteq\B^n$. Then the value of the
corresponding observable $\phi$ can be measured exactly with error probability
$\le\epsilon$\, by an operation sequence of length\,
$\poly(n)+O(n)\log(1/\epsilon)$\,
in the basis $\calQ\cup\{U^{[0,2^{2n}]}\}$. (The gate $U^{[0,2^{2n}]}$ is used
at most $O(n\,\log(n/\epsilon))$ times).
\end{theorem}

\section{Quantum algorithm for the ASP}\label{sec_qualg}

Let $(k,n,a,F)$ be an instance of the ASP,\,\,
$\St_{F}(a)=\{g\in\Z^k:\,F(g,a)=a\}$\, its solution.
Consider two finite Abelian groups\footnote{
  The group $H$ is called the group of characters on $E$.}
\begin{displaymath}
  E\ =\ \Z^k/\St_{F}(a) \qquad\qquad
  H\ =\ \mbox{Hom}(E,\T)\ \subseteq\ \mbox{Hom}(\Z^k,\T)\,=\,\T^k
\end{displaymath}
where $\T=\R/\Z$ is the group of real numbers modulo 1. Every element $h$ of
the group $H$ can be represented by $k$ rational numbers $(\bmod\ 1)$\,\,
$\phi_1,\ldots,\phi_k\in\T$\, with common denominator\, $q=|E|=|H|\le 2^n$.
More specifically, $\phi_j=(h_j,V_j)$,\, where $(\cdot,\cdot):H\times E\to\T$
is the natural bilinear mapping,\, $V_1,\ldots,V_k\in E$ are the images of the
basis elements $g_1,\ldots,g_k\in\Z^k$. It is clear that computing $\St_{F}(a)$
is polynomially equivalent to finding $H$. (To find $H$ means to find a
polynomial subset of $\T^k$ that generates this group).
We are going to show how to generate a random element of $H$
using Theorem~\ref{th_meas} (cf.~\cite{Simon}). The group $H$ itself can be
generated by sufficiently many random elements.

Consider the orbit $N=\{F(g,a):g\in\Z^k\}\,\subseteq\B^n$. Obviously, $F(g,a)$
depends only on the image of $g$ in the factor-group $E$, so we may use the
notation $g(a)$\,\, ($g\in E$). Elements of the group $E$ may be regarded as
permutations on the set $N$. The following vectors are eigenvectors for all the
operators $U\in E$
\begin{equation}
  |\psi_{h}\rangle\ =\ \frac{1}{\sqrt{q}}\, \sum_{g\in E}\,
  \exp\Bigl(2\pi i(h,g)\Bigr)\ |g(a)\rangle \qquad\qquad (h\in H)
\end{equation}
These vectors form an orthonormal basis of $\C(N)$ called the {\em Fourier
basis}. The corresponding eigenvalues are $\lambda_{h}(U)=\exp(-2\pi i(h,U))$.
In particular, if $h=(\phi_1,\ldots,\phi_k)$ then\,
$\lambda_{h}(V_j)=\exp(-2\pi i\phi_j)$\,\, ($j=1,\ldots,k$).\,
Theorem~\ref{th_meas} says that we can measure $h$ with error probability
$\le k\epsilon$\, by an operation sequence of length\,
$k\Bigl(\poly(n)+O(n)\log(1/\epsilon)\Bigl)$\,
in the basis $\calQ\cup\{G|_{O(n)}\}$.
(The operator $G$ was defined in eq.~(\ref{G})).

Now a new trick comes. Prepare the classical state
\begin{displaymath}
  |a\rangle\ =\ \frac{1}{\sqrt{q}}\, \sum_{h\in H}\, |\psi_h\rangle
\end{displaymath}
and measure $h$. By the composite probability formula (\ref{comprob}), the
probability to obtain a given value of $h$ is\,
$P(h)=q^{-1}\sum_{h'\in H}P(h',h)$,\, where $h'$ stands for the actual value
of the measured observable; $P(h',h)\ge 1-k\epsilon$. Hence
\begin{displaymath}
  q^{-1}|L|\,(1-k\epsilon)\ \le\ \Prob\,[h\in L]\ \le\ q^{-1}|L|+k\epsilon
  \qquad\quad \mbox{for any}\ L\subseteq H
\end{displaymath}
Thus we can generate a random element of H with almost uniform distribution.

Let $h_1,\ldots,h_l\in\T^k$ be independent random elements generated this way.
We are to show that they generate $H$ almost certainly, provided $l$ is large
enough.

All the elements $h_1,\ldots,h_l$ belong to $H$ with probability
$\ge 1-kl\epsilon$.
Suppose that they belong to $H$ but do not generate $H$. Then
$h_1,\ldots,h_l\in L$, where $L$ is a maximal proper subgroup of $H$. For a
given $L$, the probability of this event does not exceed\,
$\Bigl(\Prob\,[h\in L]\Bigr)^{l}\,\le\,
 \left(\frac{1}{2}+k\epsilon\right)^{l}$.\,
Maximal proper subgroups of $H$ are
in 1-to-1 correspondence with minimal nonzero subgroups of $E$. The number of
such subgroups is less than $|E|\le 2^n$. Hence the overall probability for
$h_1,\ldots,h_l$ not to generate $H$ is less than\,
$kl\epsilon+2^{n-l}(1+2k\epsilon)^{l}$.\,
(The first term corresponds to the possibility
$\{h_1,\ldots,h_l\}\not\subseteq H$ while the second one
accumulates contributions from all the subgroups $L$).
Setting\, $l=n+4$,\,\, $\epsilon=(6kl)^{-1}$\, guarantees that the random
elements $h_1,\ldots,h_l\in\T^k$ generate $H$ with probability
$\ge \frac{2}{3}$.

Thus the whole computation is organized as follows. We take $l=n+4$ registers
and prepare the initial state $|a\rangle$ in each of them. Then we do
$O(kn\,\log(kn))$ elementary measurements $\Xi\Bigl(V_j^{2^s}\Bigr)$\,\,
($1\le s\le 2n$)\, with each register. The results of these measurements are
processed in a classical way, which gives $h_1,\ldots,h_l$ and, eventually, the
canonical basis of the stabilizer (with error probability $\le\frac{1}{3}$).
Through this computation, the blackbox subroutine $F$ is invoked
$O(kn^2\log(kn))$ times for inputs of size $O(n)$. We emphasize that our
procedure is uniform, meaning that not only the operation sequence has length
$\poly(k+n)$ but also it can be constructed in time $\poly(k+n)$ by a classical
Turing machine.

\section{How to make quantum computation reversible?}\label{sec_qurevers}

In this section we will show that any quantum computation or quantum
measurement can be performed reversibly, i.e. without producing garbage.
This allows to use quantum algorithms as subroutines for other algorithms
in a non-classical way. In particular, the eigenvalue measurement procedure can
be used for the quantum Fourier transform (QFT).

We start with generalizing the definition of quantum computation
(Definition~\ref{def_quco}). Let $\Omega$ and $\Theta$ be families of mutually
orthogonal subspaces in $\C(\B^n)$ and $\C(\B^m)$, respectively. We are
going to define quantum computation for functions of type
$F:\Omega\to\Theta$. As usually, computer's memory $\Delta$ contains an input
register $X$ of size $n$ and an output register $Y$ of size $m$. Elements\,
$\calV\in\Omega$,\,\, $\calW\in\Theta$\, may be regarded as linear subspaces of
$\C(\B^X)$ and $\C(\B^Y)$, respectively. We will not make distinctions between
$\calV$ and $\calV\otimes(|0_{\Delta\backslash X}\rangle)$,\, as well as
between $\calW$ and $\calW\otimes\C(\B^{\Delta\backslash Y})$. In other words,
all the bits from $\Delta\backslash X$ are initially set to $0$, while all the
bits from $\Delta\backslash Y$ are ignored in the end.

\begin{definition} \label{def_quco1}
A unitary operator $U\in\U(\B^\Delta)$ (usually represented by an operation
sequence) is said to compute a function $F:\Omega\to\Theta$ with error
probability\, $\le\epsilon$\, if
\begin{displaymath}
  \forall\ |\xi\rangle\in\calV\in\Omega\qquad
  P\Bigl( U|\xi\rangle,\,F(\calV) \Bigr)\ \ge\ 1-\epsilon
\end{displaymath}
\end{definition}

Now we are in position to formulate an extension of
Lemma~\ref{lem_controlchange} which itself can be viewed as a
generalization of Lemma~\ref{lem_garbrem1}. In the above setting, let\,
$T=\sum_{\calW\in\Theta}\Pi_{\calW}T_{\calW}$\,
be a measurement operator for the
observable $z_{\Theta}$ with an additional register $D$. Consider the
operator\, $F_T=\sum_{\calV\in\Omega}\Pi_{\calV}T_{F(\calV)}$\,
acting on the space $\calN\otimes\C(\B^D)$, where
$\calN=\bigoplus_{\calV\in\Omega}\calV$. This is a measurement operator for
the observable $z_{\Omega}$. For applications, it is enough to consider
functions of type $F:\Omega\to\B^m$ and take the operator $\tau_m$ for $T$.
In this case the operator $F_T=F_{\tau_m}$ (or simply $F_{\tau}$) measures the
value of the function $F$ without producing any garbage. Note that for
classical functions $F$ (of type $N\to\B^m$, where $N\subseteq\B^n$)\,
the notation $F_\tau$ coincides with the notation from Sec.~\ref{sec_revers}.

\begin{theorem} \label{th_qurevers}
Let a unitary operator $U$ compute a function $F:\Omega\to\Theta$ with error
probability\, $\le\epsilon$. Let also $T$ be a measurement operator for the
observable $z_\Theta$. Then the operator $U^{-1}TU$ represents the operator
$F_T$ with precision\, $2\Bigl(|\Omega|\epsilon\Bigr)^{1/2}$.
\end{theorem}

\begin{proof}
Let $|\xi\rangle\in\calN\otimes\C(\B^D)$ be a unit vector. It can be
represented as $\sum_{\calV\in\Omega}c_\calV|\xi_\calV\rangle$,\, where
$|\xi_\calV\rangle\in\calV\otimes\C(\B^D)$ are unit vectors,\, $c_\calV\ge 0$
are real numbers. Note that $\sum_{\calV\in\Omega}c_\calV^2=1$,\, hence\,
$\sum_{\calV\in\Omega}c_\calV\le\|\Omega|^{1/2}$.\, Represent each vector
$U|\xi_\calV\rangle$ as $|\zeta_\calV\rangle+|\psi_\calV\rangle$,\, where\,
$|\zeta_\calV\rangle=\Pi_{F(\calV)}U|\xi_\calV\rangle\in
 F(\calV)\otimes\C(\B^D)$.\,
Then\,
$\langle\zeta_\calV|\zeta_\calV\rangle\,=\,
 P\Bigl(U|\xi_\calV\rangle,\,F(\calV)\Bigr)\,\ge\,1-\epsilon$,\,\,
hence\, $\|\psi_\calV\|\le\sqrt{\epsilon}$.

By the definition of the operator $T$,\,\,
$T|\zeta_\calV\rangle=T_{F(\calV)}|\zeta_\calV\rangle$,\, hence
\begin{displaymath}
  U^{-1}TU\,|\xi_\calV\rangle\ =T_{F(\calV)}|\xi_\calV\rangle\, +\,
  U^{-1}(T-1)\,|\psi_{\calV}\rangle
\end{displaymath}
The norm of the last term does not exceed $2\sqrt{\epsilon}$. Summation over
all $\calV\in\Omega$ gives the desired result.
\end{proof}

An interesting application of this theorem is a polynomial QFT algorithm for
an arbitrary finite Abelian group $G$. W.~l.~o.~g. we may take $G$ to be a
cyclic group $\Z_q$. (Transition to a direct product of cyclic groups is
straightforward). Let $q\le 2^n$, where $n$ is a constant. We identify $\Z_q$
with the set $\{0,\ldots,q-1\}\subseteq\B^n$. Our purpose is to represent the
QFT operator $V_q\in\U(\{0,\ldots,q-1\})$
\begin{equation}
  V_q|a\rangle\ =\ |\psi_{q,a}\rangle\ =\
  \frac{1}{\sqrt{q}}\, \sum_{b=0}^{q-1}\,
  \exp\left(2\pi i\frac{ab}{q}\right)\, |b\rangle
\end{equation}
by an operation sequence in the basis $\calQ$. We can also consider $q$ as a
control parameter and construct a representation for the operator
$V:\,|q,\xi\rangle\,\mapsto\,|q\rangle\otimes V_q|\xi\rangle$.

The vectors $|\psi_{q,a}\rangle$ are eigenvectors of the cyclic permutation\,
$|a\rangle\,\mapsto\,|(a+1)\,\bmod\,q\rangle$. The corresponding eigenvalues
are $\lambda_{q,a}=\exp\Bigl(-2\pi i(a/q)\Bigr)$. By Theorem~\ref{th_meas},
we can measure the value of $a$. Theorem~\ref{th_qurevers} allows us to
perform this measurement reversibly, that is to represent the following
partial operator on $\C(\B^n\times\B^n)$
\begin{equation}
  Q_q\,|\psi_{q,a},0\rangle\ =\ |\psi_{q,a},a\rangle\qquad\qquad
  (a=0,\ldots,q-1)
\end{equation}
The QFT operator $V_q$ can be constructed from the operator $Q_q$ and another
operator $T_q$ which creates the vector $|\psi_{q,a}\rangle$ for a given value
of $a$
\begin{equation}
  T_q\,|a,0\rangle\ =\ |a,\,\psi_{q,a}\rangle\qquad\qquad
  (a=0,\ldots,q-1)
\end{equation}
This construction is quite similar to that used in the proof of
Lemma~\ref{lem_garbrem2}. Let $X$ and $Y$ be two disjoint registers of size
$n$. Then\footnote{
  Recall that $\omega$ is the partial operator which maps the vector
  $|0\rangle$ to itself.}
\begin{equation}
  V_q[X]\,\otimes\,\omega[Y]\,\ =\,\
  \Bigl(Q_q[X,Y]\Bigr)^{-1}\ T_q[Y,X]\,\ \tau_n[Y,X]\,\ \tau_n[X,Y]
\end{equation}
Indeed,\,\,
$|a,0\rangle\,\mapsto\, |a,a\rangle\,\mapsto\, |0,a\rangle\,\mapsto\,
 |\psi_{q,a},a\rangle\,\mapsto\, |\psi_{q,a},0\rangle$.

It is obvious that $T_q|a,0\rangle=U_q|a,\psi_{q,0}\rangle$,\, where
\begin{displaymath}
   U_q\,|a,b\rangle\ =\ \exp\Bigl(2\pi i(ab/q)\Bigr)\, |a,b\rangle
\end{displaymath}
The operator $U_q$ can be easily constructed from
$\Lambda\Bigl(e^{2\pi i\,2^{s}/q}\Bigr)$,\, with $s=0,\ldots,n-1$\,
(by Lemma~\ref{lem_controlchange}). Thus the only remaining task is to
create the vector $|\psi_{q,0}\rangle$. For this, we have to
regard $q$ as a variable. Our procedure is recursive. For simplicity,
assume that $2^{n-1}<q<2^{n}$. At the first step, the machine sets the first
bit to the quantum state\, $(q_0/q)^{1/2}|0\rangle+(q_1/q)^{1/2}|1\rangle$,\,
where\, $q_0=2^{n-1}$,\,\, $q_1=q-q_0$. Then it looks at the value $x$
of this bit and creates the vector $|\psi_{q_x,0}\rangle$ in the
remaning $n-1$ bits. The result will be equal to $|\psi_{q,0}\rangle$.
\smallskip

\begin{remark}{Acknowledgements}
This work was supported, in part, by the ISF grant M5R000. I am grateful to
Sergei Tarasov for useful remarks.
\end{remark}


\begin{thebibliography}{99}

\bibitem{Benioff} P.~Benioff, ``Quantum mechanical Hamiltonian
models of Turing machines", {\it J.~Stat.~Phys.} {\bf 29}, 515 (1982).

\bibitem{Peres} A.~Peres, ``Reversible logic and quantum computers",
{\it Phys.~Rev.} {\bf A 32}, 3266 (1985).

\bibitem{Feynmann} R.~P.~Feynman, ``Quantum mechanical computers",
Optics News, February 1985, {\bf 11}, p.~11.

\bibitem{Deutsch1} D.~Deutsch, ``Quantum theory, the Church-Turing principle
and the universal quantum computer",
{\it Proc.~R.~Soc.~Lond.} {\bf A 400}, 97 (1985).

\bibitem{Deutsch2} D.~Deutsch, ``Quantum computational networks",
{\it Proc.~Roy.~Soc.~Lond.} {\bf A 425}, 73 (1989).

\bibitem{Yao} A.~C.-C.~Yao, ``Quantum Circuit Complexity",
{\it Proceedings of the 34th Annual Symposium on the Foundations of Computer
Science} (IEEE Computer Society Press, Los Alamitos, CA, 1993), p.~352.

\bibitem{Shor} P.~W.~Shor, ``Algorithms for quantum computation:
discrete log and factoring",
{\it Proceedings of the 35th Annual Symposium on the Foundations of
Computer Science} (IEEE Computer Society Press, Los Alamitos, CA, 1994),
p.~124.

\bibitem{Bernstein-Vazirani} E.~Bernstein and U.~Vazirani,
``Quantum complexity theory'',
{\it Proceedings of the 25th Annual ACM Symposium on Theory of Computing},
(ACM Press, New York, 1993), \mbox{pp.~11\,--\,20}.

\bibitem{Grigoriev} D.~Grigoriev, ``Testing shift equivalence of polynomials
using quantum machines'' {\it (to appear)}

\bibitem{Simon} D.~Simon, ``On the power of quantum computation",
{\it Proceedings of the 35th Annual Symposium on the Foundations of
Computer Science} (IEEE Computer Society Press, Los Alamitos, CA, 1994),
p.~116.

\bibitem{Deutsch-Jozsa}
D.~Deutsch, and R.~Jozsa,
``Rapid solution of problems by quantum computation'',
{\it Proceedings of the Royal Society}, London,
{\bf A439}, 1992, 553--558.

\bibitem{Coppersmith}
D.~Coppersmith, ``An approximate Fourier transform useful in quantum
factoring", {\it IBM Research Report RC19642} (1994)

\bibitem{factoring}
G.~L.~Miller, ``Riemann's hypothesis and tests for primarity'',
{\it J.~Comp.~Sys.~Sci}, {\bf 13}, 300--317 (1976)

\bibitem{Lecerf}
Yves Lecerf, ``Machines de Turing reversibles.
Recursive insolubilite en $n \epsilon N$ de l'equation $u=\theta^n$ ou
$\theta$ est un ``isomorphism de codes''. {\em Comptes Rendus}
{\bf 257}, 2597--2600 (1963)

\bibitem{Bennett}
C.~H.~Bennett, ``Logical reversibility of computation",
{\it IBM Journal of Research and Development} {\bf 17}, 525 (1973).

\bibitem{9authors}
A.~Barenco, C.~H.~Bennett, R.~Cleve, D.~P.~DiVincenzo, N.~Margolus,
P.~Shor, T.~Sleator, J.~Smolin, and H.~Weinfurter,
``Elementary gates for quantum computation'', \mbox{quant-ph/9503016}

\end{thebibliography}
\end{document}